\documentclass[twocolumn,showpacs,amsmath,amssymb,floatfix]{revtex4}
\usepackage{graphicx}
\usepackage{dcolumn}
\usepackage{bm}

\input epsf
\draft

\begin{document}

\title{Equation of State of Supercooled
Water\\ from the Sedimentation Profile} 

\author{Masako Yamada,$^1$ H. Eugene Stanley$^1$ and Francesco Sciortino$^2$}

\affiliation{
$^1$Center for Polymer Studies and Department of Physics,
Boston University, Boston, MA 02215 USA \\
$^2$Dipartimento di Fisica, INFM UdR and INFM Center for
Statistical Mechanics and Complexity,
Universit\`a di Roma ``La Sapienza'', Piazzale Aldo
Moro 2, I-00185, Roma, Italy\\
}

\date{yss.tex printed \today}

\begin{abstract}

To study the coexistence of two liquid states of water within one
simulation box, we implement an equilibrium sedimentation method---which
involves applying a gravitational field to the system and
measuring/calculating the resulting density profile in equilibrium.  We
simulate a system of particles interacting via the ST2 potential, a model
for water. We detect the coexistence of two liquid phases at low temperature.

\end{abstract}

\pacs{PACS number(s):}
\maketitle


\section{Introduction}

The physics of the liquid state has been the subject of intense research
activity.  Novel approaches and novel apparatus have made possible the
study of liquids under extreme thermodynamic conditions and in wide
windows of space and time.  Interesting and unexpected new phenomena have
emerged as a result of a combined effort involving experiments, theory
and simulations\cite{Debenedetti, MarieClaire}.  One of these is the 
possibility of a liquid-liquid (LL) transition in one-component 
systems---in addition to the usual liquid-gas transition. Several liquids 
\cite{Katayama, Ree, Togaya} and several models 
\cite{StellHemmer,GF,Reza,Swendsen,Jagla,Pablo,Saika,Lacks} 
have been studied in detail, and it appears that the class of materials 
where a LL transition can be observed is larger than the class of 
tetrahedral liquids that were originally considered \cite{Angell00} 
possible candidates for a LL transition.

Water is one of the liquids that might possess a LL transition.  Indeed,
the first conjecture of a LL transition was based on a numerical study
of the ST2 potential \cite{Poole}, a model designed to mimic the
behavior of liquid water.  In the case of ST2, the two coexisting phases
differ in their local structure. The low density phase is formed by an
open tetrahedrally-coordinated network of hydrogen bonds, while
the high density phase has a more distorted network of hydrogen bonds. 
Recent theoretical work has shown that the interplay between
local energy, entropy, and volume which may generate a LL transition can
in principle be realized by spherically-symmetric potentials
\cite{GF,Jagla}.

The evaluation of the $P(\rho,T)$ equation of state (EOS) is key to test
for phase coexistence (gas-liquid, liquid-crystal and LL),
where $P$, $\rho$, and $T$ denote the pressure, density, and
temperature.  The numerical calculation of $P(\rho,T_0)$ requires the
study of the model for a variety of state points.  The coexistence
between two phases, in the appropriate temperature window, appears as a
region of density values where $P$ is constant. In small-size numerical 
simulations is sometimes hard to observe phase transitions directly in one 
simulation box, in part because the free energy associated with creating 
an interface often stabilizes metastable phases\cite{Panagiotopoulous}. In 
these cases $P(\rho,T)$ does not show any flat region \cite{negative}.

A different approach for studying in one single numerical simulation an
entire isotherm has been proposed in Ref.~\cite{Barrat} (and later
exploited in the experimental study of colloidal systems and in the study
of crystallization profiles Ref.~\cite{Piazza, Lowen, Allain, Chaikin}).  
This approach simulates a semi-infinite tube in the presence of a very
strong gravitational field and measures the density profile in
equilibrium.  A simple inversion of the density profile allows the model
EOS to be constructed. This idea has also been applied in the experimental
study of the EOS of colloidal particles, by inversion of the measured
sedimentation equilibrium profile.

Here we apply the sedimentation profile method to ST2 water. We find that
an interface separating two liquid states appears at low temperature,
corresponding to the coexistence of two metastable liquid states of
water within one simulation box, providing  evidence for the
presence of an LL transition.

\section{Theory and Simulation Details}

We study $N=7680$ rigid molecules of mass $M$ interacting through the
ST2 water potential, a rigid, non-polarizable, 5-site potential
\cite{Stillinger} that is able to reproduce qualitatively the
thermodynamic anomalies of liquid water.  ``ST2 water'' is
characterized, on cooling, by isobaric density maxima, increasing
compressibility, increasing constant $P$ specific heat, and evidence for
a LL transition in the deeply supercooled regime (which
is difficult to probe experimentally due to spontaneous crystallization
\cite{Poole}).

\begin{figure} 
{\epsfxsize=5cm \epsfbox{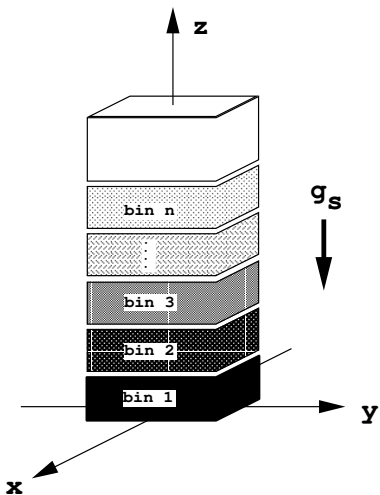}} 
\caption{Schematic of the columnar box in a gravitational field
$g_s$. We partition the box into equally sized bins
with height $\Delta z$, where each bin represents a state point. The
number of molecules in each bin is used to calculate the density and
pressure at the center of the bin.}
\label{fig_column} 
\end{figure}

To implement the sedimentation profile method, we use a
column-shaped simulation box, semi-infinite along the $z$ axis and with
periodic boundary conditions along the $x$ and $y$ axes. The top of the
box is left open, while the bottom is assumed to be a repulsive
soft-sphere surface, generating a short-range force proportional to
$z^{-13}$ acting on the molecule center of mass (Fig.~\ref{fig_column}). 
The box width in the $x$-axis and $y$-axis directions is 3~nm, 
corresponding to a bottom surface area of $S_{xy}=9$~nm$^2$.  A strong
gravitational field $g_s=2\times 10^{12}g$ is applied downward, in the 
$z$-axis direction, where $g=9.8$~kg/m$^2$ is the Earth's gravitational
field.  The value of the field controls the range of 
$P$-values accessed in the simulations. The pressure at the bottom of 
the column is $P=g_sNM/S_{xy} \simeq 500$MPa.

We perform the simulations using a multi-processor code on SGI Origin
2000, IBM~SP, and IBM Regatta supercomputers. We choose a 1~fs time step
and study different temperatures from $T=300$~K down to $T=230$~K. 
For T=$230$~K, we simulate four different systems to better estimate error 
and reproducibility of the results. A long ($\sim 10ns$) equilibration time
precedes the actual calculation of the equilibrium density profile.  To
analyze equilibration we monitored the running average of the
$z$-position of the center of mass.  Production runs lasted at least
200~ps for the higher temperatures up to several ns per box when 
$T=230$~K.


We define a mass density field $\rho(z)$ by averaging the density over
bins with height $\Delta z=1$~nm to calculate the pressure field $P(z)$
from
\begin{equation}
P(z)=  g \int_z^\infty\rho(z')dz'.
\label{pres_eq2}
\end{equation}
A parametric plot of $P(z)$ vs. $\rho(z)$ provides the EOS.

\section{Results and Discussion}

\begin{figure} {\epsfxsize=7.5cm \epsfbox{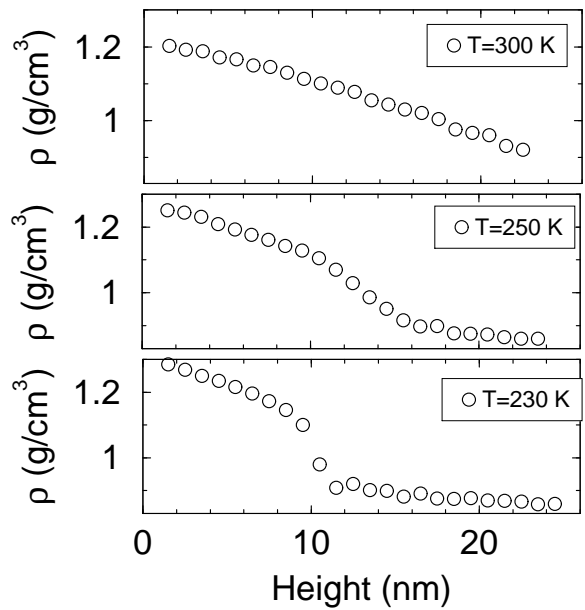}} 
\caption{Density profiles for ``ST2 water'' for three different
temperatures. At T=230 K a discontinuity in the density 
appears around 10 nm. In all boxes, the liquid-gas transition is at the
state point at the maximum height shown.}
\label{fig:height} 
\end{figure}

Figure~\ref{fig:height} shows the equilibrium density profile along the 
$z$-axis for three different temperatures. Each symbol represents one bin, 
where the height of each bin $\Delta z=1$~nm. For $T=300$~K we see only 
one break in the density profile: the topmost points in the plots 
correspond to the gas-liquid interface. Above these points is a much less 
dense gas not shown on this density scale. For $T=250$~K, there appears an
inflection. For $T=230$~K, we see a clear break in the density,
associated with the interface between two different liquid states.

\begin{figure} {\epsfxsize=7.5cm \epsfbox{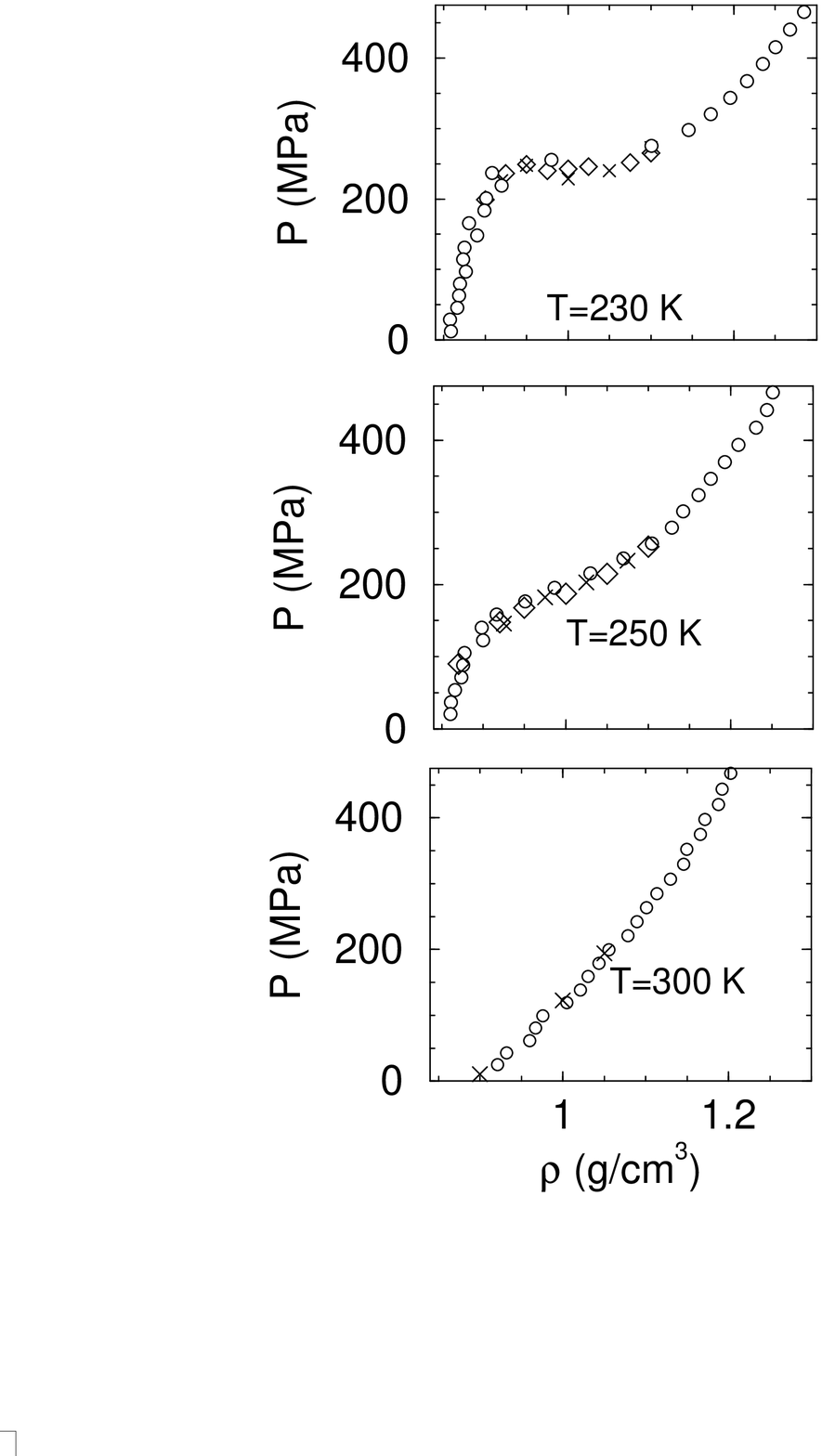}} 
\caption{EOS for the ST2 potential using the sedimentation method shown
as open circles. Open
diamonds are the $P(\rho,T)$ previously reported
in Ref.~\protect\cite{peter}, and the crosses are values
previously reported in Ref.~\protect\cite{steve}. We corrected
the data of Ref.~\protect\cite{peter} and Ref.~\protect\cite{steve}
to subtract the correction to $P$ arising from  the integration
of the Lennard-Jones potential beyond the cut-off implemented
in the simulation.  Note the unphysical ``loop'' in the EOS
at $T=230 K$ which can be observed in the standard cubic-box simulations.
}

\label{fig:eos} 
\end{figure}

Figure~\ref{fig:eos} shows the corresponding $P(\rho)$ relations at the
same three temperatures. We also show $P(\rho)$ as evaluated previously
using standard MD for cubic boxes with periodic boundary conditions for
systems with $N=6^3=216$ (Ref.~\cite{peter}) and $N=12^3=1768$ 
(Ref.\cite{steve}). In Refs~\cite{peter} and \cite{steve} $P$ was 
calculated with the standard virial relation\cite{negative2}.

Consistent with the data shown in Fig.~\ref{fig:height}, we see a region
of coexistence between two different phases at $T=230 K$, in
agreement with the estimated location of the LL transition in
Ref.~\cite{Sciortino}.  Note that when using the sedimentation
equilibration method, we see no unphysical loops in the equation of
state, unlike cubic box simulations where the boundary conditions may
artificially stabilize metastable states.

\begin{figure} {\epsfxsize=7.5cm \epsfbox{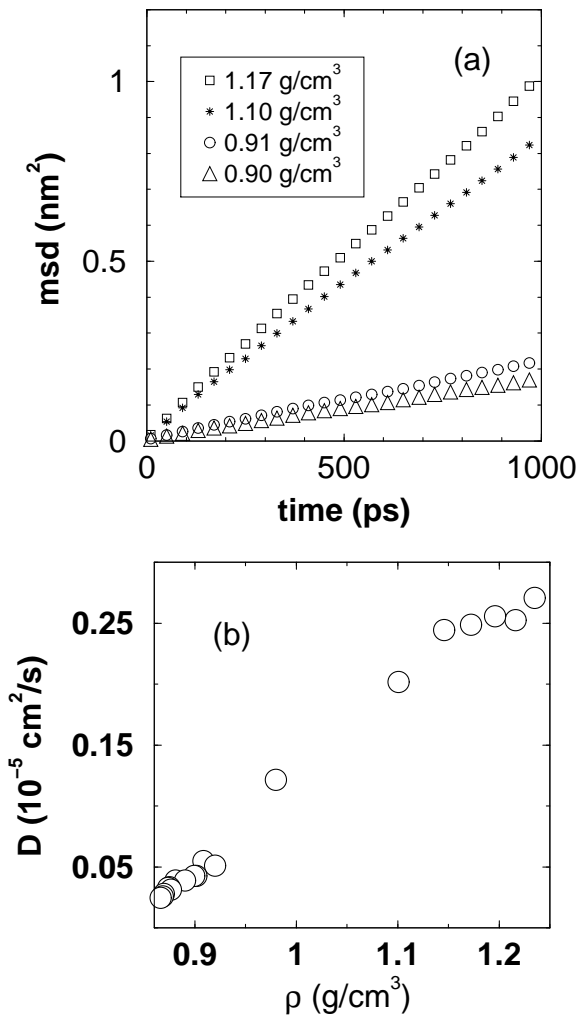}} 
\caption{(a) Mean square displacement of molecules for different density 
values within the column-shaped simulation box.  
Note that for values of the density both lower and
higher than the coexisting phases, molecules diffuse with a finite
diffusion constant. (b) Dependence 
of diffusion constant on density $\rho$ across the interface.}
\label{fig:diff} 
\end{figure}

To confirm that the two coexisting phases are both liquids, we calculate
the mean square displacement (MSD) for different height values, both
below and above the interface.  We follow the evolution of each molecule
for an average mean squared displacement smaller that $\Delta z$, so
that each height value can be unambiguously assigned to an average
density value. Figure~\ref{fig:diff} shows that both phases are
sufficiently diffusive. The low density phase has smaller diffusivity, in
agreement with previous simulations of the density dependence of the
dynamics \cite{Geiger}.

\section{Conclusions}

We studied the EOS of ``ST2 water'' with the sedimentation profile method,
and presented evidence that the EOS at $T=230 K$ shows a clear phase
coexistence, between two phases which are both liquids.  The observed LL
coexistence in the ST2 potential phase diagram is consistent with estimate
of the ``critical point'' being located around $T=235$~K, $P=250$~MPa, and
$\rho=1.05$~g/cm$^3$. Note that for most simple water models, the
temperature and pressure scales are shifted relative to real values, thus
they must be shifted to place them within an experimental
context\cite{compare}. The critical point estimated experimentally is
around $T_c=230$~K and $P_c=50$~MPa with
$\rho=1.05$~g/cm$^3$\cite{Mishima}.

\section*{Acknowledgments}

This work is supported by NSF grant CHE-0096892. MY is also supported
by NSF grant GER-9452651. FS thanks INFM-Iniziative Parallel
Computing for granting numerical resources.



\begin{references}
\bibitem{negative}

In small-size simulation boxes, a negative slope ``loop'' may be seen, and
the Maxwell construction method is used to calculate the critical pressure
and density..
\bibitem{negative2}
Regions of negative $P$ cannot be studied with the sedimentation method, 
at least not in its present form.
\bibitem{compare}

In comparison, the critical point estimated using the using the SPC/E
potential is estimated to be $T=130$~K and $P=290$~MPa with
$\rho=1.10$~g/cm$^3$ \cite{Scala}; using the TIP4P potential $T<200$~K and
$P_c>70$~MPa \cite{Sciortino}; and using the TIP5P potential $T_c\approx
217$~K, $\rho_c\approx 1.13$~g/cm$^3$, and $P\approx 340$~MPa \cite{TTT}.
\end{references}
\end{document}